\documentclass[aps,pre,twocolumn,superscriptaddress,floatfix]{revtex4}
\usepackage{graphicx}
\usepackage[dvips,unicode,colorlinks,linkcolor=blue,citecolor=blue,urlcolor=blue]{hyperref}
\usepackage{times}

\begin{document}
\title{Heat conduction in a chain of dissociating particles: effect of dimensionality}
\author{V. Zolotarevskiy}
\affiliation{
Faculty of Mechanical Engineering, Technion -- Israel Institute of Technology, Haifa 32000, Israel}

\author{A. V. Savin}
\affiliation{Semenov Institute of Chemical Physics, Russian Academy of Sciences, Moscow 119991, Russia}

\author{O. V. Gendelman}
\email{ovgend@tx.technion.ac.il}
\affiliation{
Faculty of Mechanical Engineering, Technion -- Israel Institute of Technology,
Haifa 32000, Israel}

\date{\today}

\begin{abstract}
The paper considers heat conduction in a model chain of composite particles with hard core and
elastic external shell. Such model mimics three main features of realistic interatomic potentials
-- hard repulsive core, quasilinear behavior in a ground state and possibility of dissociation. It has
become clear recently, that this latter feature has crucial effect on convergence of the heat
conduction coefficient in thermodynamic limit. We demonstrate that in one-dimensional chain
of elastic particles with hard core the heat conduction coefficient also converges, as one could expect.
Then we explore effect of dimensionality on the heat transport in this model. For this sake,
longitudinal and transversal motions of the particles are allowed in a long narrow channel.
With varying width of the channel, we observe sharp transition from "one-dimensional"\ to "two-dimensional"\
behavior. Namely, the heat conduction coefficient drops by about order of magnitude for relatively
small widening of the channel. This transition is not unique for the considered system.
Similar phenomenon of transition to quasi-1D behavior with growth of aspect ratio of the channel
is observed also in a gas of densely packed hard (billiard) particles, both for two- and
three-dimensional cases. It is the case despite the fact that the character of transition in these two systems is not similar,
due to different convergence properties of the heat conductivity.
In the billiard model, the divergence of the heat conduction coefficient smoothly changes
from logarithmic to power-like law with increase of the length.
\end{abstract}

\maketitle

\section{Introduction}
Fourier law of heat conduction has remained for two hundred years one of the most important
topics in thermal physics. Empiric results show accurate validation of Fourier
proposition. In the same time,  relationship between equations of heat conduction and
microstructure of solid dielectrics is known to be one of the oldest and most elusive unsolved
problems in solid state physics, with considerable research efforts over last
three decades \cite{FPU}-\cite{AA99}.

Significant step in the study of heat transport was carried out in seminal numerical
experiment by Fermi, Pasta and Ulam (FPU) in 1954 \cite{FPU}. The idea was to show
that a simple one-dimensional system can acquire statistical-mechanical properties independently
of the initial conditions. They presented the crystal as a one-dimensional chain of equal
oscillators with nearest-neighbor interaction, with potential including quadratic as well as cubic
and quartic terms. FPU assumed that the dynamic evolution will eventually lead to energy
equipartition between all the linear modes of the system, as the thermal equilibrium is
established. However, the system did not show the expected behavior. The energy was
exchanged only among the lowest modes, and then restored to nearly initial configuration.
The result disproved a common belief on inevitable fast thermalization and mixing in non-integrable
systems with weak nonlinearity.

The anomaly of thermalization, which was observed by FPU, is not unique to the system studied
in their experiment. The most well-known examples of such anomalies are one-dimensional integrable
systems, such as harmonic and Toda lattices, in which the heat flux does not depend on the system size,
but rather on the temperature difference. Consequently, such systems have a divergent heat
conductivity coefficient, as the length of the system increases \cite{LLP08,Ht}.
Moreover, even linear temperature distribution is never established in these integrable systems.

Over recent years, numerous additional anomalies in the heat transfer in microscopic models of
dielectrics were revealed by means of direct numeric simulation, including qualitatively different
behavior of models of different types (with and without on-site potential) \cite{LLP03}.
It is widely believed (with some counter-examples discussed below) that in one dimension the heat
conduction coefficient in the microscopic models with conserved momentum diverges in the
thermodynamic limit (as the chain length $N$ goes to infinity) as $\kappa\sim N^\beta$
with $\beta$ varying in the interval $0.3\div 0.4$ \cite{LLP03, LiR}.

Recent work on 1D chain of semi-elastic rods, as well as on more traditional models with
Lennard-Jones and Morse potential showed convergence of heat conductivity coefficient \cite{SK14,GS14}.
Divergent heat conduction of isolated low-dimensional systems can be explained by a weak scattering
of long-wavelength phonons, which possess long mean free paths.
Then, it is possible to conjecture that finite
conductivity has to be related to some well-defined mechanism, which enables efficient phonon scattering.
So, convergence of the heat conduction coefficient in one-dimensional models was observed due to specific
choice of boundaries \cite{RD}, in the chain of coupled rotators \cite{GLPV,GS00,GS01},
and recently in chains capable of dissociation \cite{SK14, GS14}. In this latter case
the thermally activated "gaps"\  in the chain ensure efficient phonon scattering,
sufficient for the convergence of the heat conduction coefficient.

Two-dimensional system of anharmonically interacting oscillators with conserved momentum is
also expected to have divergent heat conductivity.
In particular, a logarithmic divergence of the conductivity with system size is
predicted by mode coupling theory \cite{LLP08,LLP03,LiR,D}. The first numerical study on heat
transport problem of two-dimensional lattice was presented in work of Payton and Visscher \cite{PV}.
Dependence of the heat conductivity coefficient on the size of the system was considered
in later work by Jackson and Mistriotis \cite{GM} that conducted a comparison of 1D
and 2D FPU lattices; infinite conductivity has been observed. Recent explorations of the 2D systems
predict anomalous heat conduction, with either logarithmic
(see \cite{LL} for FPU and Lennard-Jones lattice), or power-law
(see \cite{GY,SI} for FPU) divergence of the heat conductivity coefficient.
The three-dimensional case arises a lot of controversy \cite{SI,SD}.

All studies mentioned above were aimed at exploration of lattices infinite in all directions
(within obvious numeric restrictions). Different, and quite interesting, situation arises, when
some dimensions exist in the model, but are externally confined. Besides purely academic
interest, such models might be useful for understanding the thermal behavior of nanosystems with
large aspect ratio. We are going to concentrate on "quasi-one-dimensional"\ models, in which
only one dimension is spatially extended, and the thermodynamic limit is considered only in one
direction. Deutsch and Narayan \cite{DN03} studied the thermal conductivity of such quasi-1D
chains of hard spheres. In this model, the spheres had an additional degree of freedom, but
a modification of the initial order was not allowed. They found that the conductivity of a system
of spheres with equal masses, and also with alternating masses, diverges with a size of the system.
Similar results were obtained by Lipowski and Lipowska \cite{LL07}, for quasi-1D models of hard
disks that can and cannot exchange their positions. Anomalous heat conductivity
was also seen in the recent work of Morriss and Truant \cite{MT13} on non-interacting
hard disks in a channel.

An interesting idea for simulating the finite conductivity was to introduce scatterers
in quasi-1D billiard gas channels. The first work in this field was carried out
by Alonzo et al. \cite{AA99}, where a quasi-1D billiard in Lorentz gas channel was analyzed.
The ends of the channel were inserted into heat baths, and the movement of the particles inside
the channel was interrupted by semicircular scatterers. The conductivity
of such a chaotic system obeyed Fourier's law. In order to investigate
the role of chaos on the problem of heat conduction, following works implemented changes in
geometry and order of the scatterers \cite{LiR}. Some of the "modified" configurations showed
normal heat conductivity. In some others, the heat conduction coefficient diverged \cite{LiR}.
Thus the assumption that chaos may be sufficient condition for a system to possess the finite
heat conductivity has been disproved. Although such billiard gas models might shed some light
on heat transfer, they lack particle interaction, phonon transport, and local thermal equilibrium,
and cannot represent the oscillatory lattice-like structure \cite{LiR}.

Previous studies considered models with only one possible mechanism of the heat transport:
oscillatory waves in a system with fixed microstructure or moving particles of\ "rarefied gas"\ in
the channel without collisions.
However, it is easy to imagine physical situation, in which these mechanisms will
co-exist. For instance, one can consider dense gas in closed channel with two of the walls acting as
thermostats. One can expect that main mechanism of the transport will be still related to wave propagation.
From the other side, individual particles can move separately and even exchange their positions,
and so the transport through the motion of individual particles is also possible.
Two particular cases mentioned above (crystal and non-interacting particles in a channel) are
natural limits of such model for very high and very low densities respectively. The goal of
present paper is to explore the heat conduction in quasi-1D chain with two competing mechanisms
of heat transport.

In such simulation it is desirable to exclude effects related to anomalies of the heat transport
in low-dimensional systems. In order to achieve that, according to \cite{GS14},
we consider here the model with possibility of complete dissociation. In order to make the model
closer to physical reality, we also include a repulsive hard core. So, the considered model
consists of particles with hard core covered by deformable shell. So, both phonon-like oscillatory
waves and individual motion of the particles are possible. The heat transport is simulated
in a chain of such particles imbedded into a channel with rigid walls.
In this model each particle can move in two dimensions. From the other side, the geometry of
the channel dictates large aspect ratio and therefore one can state that we consider
"quasi-one-dimensional"\  model. Our main goal is to probe this "quasi-one-dimensionality"\
through variation of the channel width. For the sake of comparison, we simulate also a chain
of "billiard"\  particles in the same quasi-1D setting with the possibility of 2D and 3D motion.

\section{Description of the model}
Let us consider the one-dimensional chain which consists of $N$ disks with elastic compressive
interaction. The diameter of the disks is $D>0$, and the disks have a hard core with diameter $D_0$,
$0<D_0<D$. The disks repulse each other, if their centers are at a distance less then $D$.
The potential of interaction is defined as:
\begin{eqnarray}
U(R)=\infty,~~\mbox{for}~~R\le D_0, \nonumber\\
U(R)=\frac12 K(D-D_0)^2 \left(\frac{D-R}{R-D_0}\right)^2,~\mbox{for}~D_0<R\le D, \label{f1} \\
U(R)=0,~~~\mbox{for}~~R\ge D, \nonumber
\end{eqnarray}
where $R$ is the distance between the centers and $K=U''(D)>0$ characterizes the stiffness
of the disks. The potential $U(R)$ vanishes at the distance
$R \ge D$, increases monotonically as $R$ decreases, and approaches infinity when $R\rightarrow D_0$.
We may notice that for $D_0=0$ the expression (\ref{f1})
takes a common form of Lennard-Jones 1-2 potential.

The Hamiltonian of the chain takes the following form
\begin{equation}
H=\sum_{n=1}^N \frac12M(\dot{\bf R}_n,\dot{\bf R}_n)+\sum_{n=1}^{N-1}\sum_{m=n+1}^N
U(|{\bf R}_m-{\bf R}_n|),\label{f2}
\end{equation}
where $M$ is the mass of each disk and ${\bf R}_n$ stands for the position of the $n$-th disk.
We introduce dimensionless displacement $r=R/D$, dimensionless energy
${\cal H}=H/KD^2$ and dimensionless time $\tau=t\sqrt{K/M}$.
orresponding dimensionless Hamiltonian is written as
\begin{equation}
{\cal H}=\sum_{n=1}^N \frac12({\bf r}'_n,{\bf r}'_n)+\sum_{n=1}^{N-1}\sum_{m=n+1}^N
V(|{\bf r}_m-{\bf r}_n|),\label{f3}
\end{equation}
where the apostrophe denotes differentiation with respect to $\tau$,
${\bf r}_n={\bf R}_n/D$ is the dimensionless position vector of $n$-th disk, $0<d=D_0/D<1$
is dimensionless core, the dimensionless repulsive interaction between disks is
\begin{eqnarray}
V(r)&=& \infty,~~\mbox{for}~~r\le d, \nonumber \\
V(r)&=&\frac12(1-d)^2\left(\frac{1-r}{r-d}\right)^2,~~\mbox{for}~~d<r< 1, \label{f4}\\
V(r)&=& 0,~~\mbox{for}~~r\ge 1. \nonumber
\end{eqnarray}
In order to be more specific, we will use the dimensionless value
$d=0.8$ for the diameter of the hard core.

\section{Heat Conduction in the one-dimensional chain}\label{s3}
We start with traditional numeric simulation of heat transport in one-dimensional model
of particles described in the previous section. It is easy to notice, that  potential (\ref{f4})
has a discontinuity of second derivative at $r=1$.
To avoid numeric complications, we will approximate it by smoothened potential.
The smoothening procedure is described in Appendix \ref{a1}.

Let us consider a segment of length $L$ parallel to $x$ axis. We pack $N=p(L-1)+1$ disks
along this segment, where $p$ ($0<p<1/d$) stands for the packing "density"\ of the chain.
Fixed boundary conditions are imposed on both ends of the chain, i.e. $x_1=0$, $x_N\equiv (N-1)a$,
where $a=1/p$ stands for the period of the unperturbed chain. Fixed boundaries enable
the density conservation. The disks $1<n<N$ are then restricted to move in
$x$ direction. The Hamiltonian of the chain in this case is expressed as
\begin{equation}
{\cal H}=\sum_{n=2}^{N-1}\frac12{x'_n}^2+\sum_{n=1}^{N-1}V(x_{n+1}-x_{n}).
\label{f5}
\end{equation}
Here $\{x_n\}_{n=1}^N$ are coordinates of disk centers.

To model the heat transfer along the chain under consideration we will use a stochastic Langevin
thermostat.
A left end ($L_0=10$) of the chain is inserted into Langevin thermostat with temperature $T_+$,
and  the right end of the chain with the same length  -- into thermostat with temperature $T_-$.
We adopt  $T_\pm=(1\pm 0.05)T$, where $T$ is average temperature of the chain.
The corresponding equations of motion has form:
\begin{eqnarray}
x''_n&=&-\partial {\cal H}/\partial x_n -\gamma x'_n+\xi_n^+,~~\mbox{if}~~x_n< L_0,\nonumber\\
x''_n&=&-\partial {\cal H}/\partial x_n,~~\mbox{if}~~L_0\le x_n\le (N-1)a-L_0,\label{f6}\\
x''_n&=&-\partial {\cal H}/\partial x_n -\gamma x'_n+\xi_n^-,~~\mbox{if}~~x_n> (N-1)a-L_0,\nonumber
\end{eqnarray}
where  $\gamma=1/t_r$ is a damping coefficient with time of relaxation
$t_r$, $\xi_n^\pm$ is  Gaussian white noise which models the interaction with the thermostats,
and is normalized by the conditions  $\langle\xi_n^\pm(\tau)\rangle=0$,
$\langle\xi_n^+(\tau_1)\xi^-_k(\tau_2)\rangle=0$,
$\langle\xi_n^\pm(\tau_1)\xi_k^\pm(\tau_2)\rangle=2\gamma T_\pm\delta_{nk}\delta(\tau_2-\tau_1)$.

System of equations (\ref{f6}) with initial conditions
${\bf X}(0)=\{x_n(0)=(n-1)a,~x'_n(0)=0\}_{n=1}^N$ was integrated numerically by Velocity Verlet method.
The thermal equilibrium between the chain and the thermostats is then reached and is manifested
by a stationary heat flux $J$ and stationary local temperature distribution $T(x)$.

The total heat flux $J$ is defined in terms of the mean value of the work produced by the
thermostats over unit time. For this matter at each step of numerical integration $\Delta\tau$
new coordinates of the disks were calculated without account of the interaction with thermostats
${\bf X}_0(\tau+\Delta\tau)$ and then the same coordinates were calculated for chain interacting
with the thermostats, denoted as ${\bf X}(\tau+\Delta\tau)$. We define $E_+$ as the energy of the left most
segment of the chain which consists of disks with coordinates $x_n<L/2$ and $E_-$ as energy
of the right most segment, where disks have coordinates $x_n>L/2$. Then the work done by
the external forces in the time interval $[\tau,\tau+\Delta\tau]$ is
\begin{equation}
j_\pm=[E_\pm({\bf X}(\tau+\Delta\tau))-E_\pm({\bf X}_0(\tau+\Delta\tau))]/\Delta\tau.
\label{f7}
\end{equation}

By taking time average $J_\pm=\langle j_\pm\rangle_\tau$ we obtain the average value of energy
flux-out from the left "hot"\ thermostat and the average value of the energy flux-in into the
right "cold"\  thermostat. The value of energy flux along the chain is  $J=J_+=-J_-$.
Accuracy of this balance is considered as a criterion for validity of our numeric simulation.

The local heat flux, i.e. the energy flow from disk $n$ to the neighboring disk $n+1$, is defined
as $J_n=\langle j_n\rangle_\tau$, where
$$
j_n=\frac12(x_{n+1}-x_n)(x'_{n+1}+x'_n)F(x_{n+1}-x_n)+x'_nh_n,
$$
function $F(r)=-dV(r)/dr$, energy density distribution along the chain
$$
h_n=\frac12\left[{x'_n}^2+V(x_{n}-x_{n-1})+V(x_{n+1}-x_{n})\right].
$$
(see \cite{LLP03}).

The thermal equilibrium requires all local fluxes to be equal to the total heat flux multiplied
by the chain period,  $J_n=aJ$. The fulfillment of this requirement may be considered as
a criterion for stationary regime of the heat transport.

The local temperature distribution of the chain is calculated from kinetic energy of the disks.
Let us divide the line segment $L$, which consists of $N$ disks, into unit-length cells $[i-1,i]$,
$i=1,...,L$. We define the following quantities:
the average number of disks in $i$-th cell is $\bar{n}_i$, and
the average kinetic energy in the cell $\bar{E}_i$. Then the temperature of the cell $T(i)=2\bar{E}_i/\bar{n}_i$.
%---------------------------- Fig. 1 ------------------------------------
\begin{figure}[tb]
\includegraphics[angle=0, width=1\linewidth]{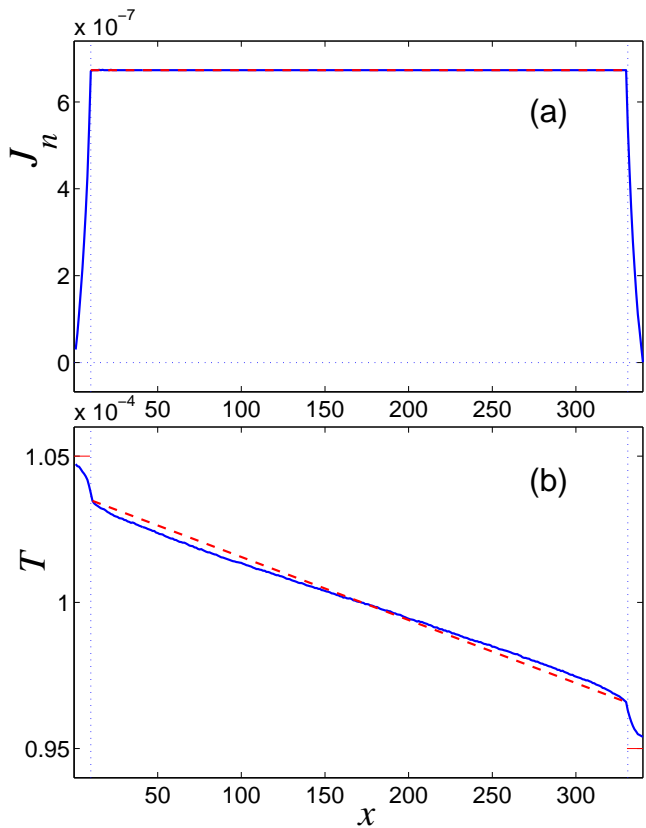}
\caption{(Color online)
Distribution of (a) the local heat flux $J_n$ and (b) of temperature $T(x)$ along the chain of length
$L=340$. The density of the disks packing is $p=1$ (average distance between disk centers is  $a=1/p=1$),
the temperature of the thermostats $T_+=0.000105$, $T_-=0.000095$.
The red horizontal dashed curve in part (a) represents the value of the heat flux  $J$.
The red dashed line in part (b) represents the linear temperature gradient.}
\label{fig01}
\end{figure}
%---------------------------- Fig. 1 ------------------------------------

A representative example of heat flux and temperature distribution in a 1D chain is presented in
Fig.~\ref{fig01}. We infer that at the internal fragment of the chain  $L_0<x<L-L_0$ the heat flux
is constant and independent of the number of disk ($J_n=aJ$) and the temperature profile is almost
linear. Then, we evaluate the heat conduction coefficient of the chain for the internal
fragment with length $\tilde{L}=L-2L_0$:
\begin{equation}
\kappa=J(L-2L_0)/[T(L_0)-T(L-L_0)].
\label{f8}
\end{equation}

The heat conduction coefficient converges in the thermodynamic limit if the following limit exists:
\begin{equation}
\bar\kappa=\lim_{L\rightarrow\infty}\kappa(L).
\label{f9}
\end{equation}

In the numeric simulation of the heat transport we considered
chain length intervals $L=20+10\times2^{k-1}$, $k=1$, 2, ..., 11.
The length of terminal segments of the chain, where it interacts
with the thermostats, was taken as $L_0=10$. The relaxation time of disk velocity was $\tau_r=10$.

The heat conductivity coefficient may be also obtained using Green-Kubo formula \cite{GK}:
\begin{equation}
\kappa_c=\lim_{\tau\rightarrow\infty}\lim_{L\rightarrow\infty}\frac{1}{LT^2}\int_0^\tau c(t)dt,
\label{f10}
\end{equation}
where $c(\tau)=\langle J_s(t)J_s(t-\tau)\rangle_t$ is an autocorrelation function of the
total heat flux in the chain
$
J_s(\tau)=\sum_{n=1}^N j_n(\tau).
$

In order to calculate the autocorrelation function $c(\tau)$ we considered a cyclic chain
consisting of $N=10^4$ particles with fixed overall length $L=N/p$. Initially all disks are
coupled to the Langevin thermostat with temperature $T$. After achieving the thermal equilibrium,
the system is detached from the thermostat and Hamiltonian dynamics is simulated.
To improve the accuracy, the results were averaged over $10^4$ realizations of the initial thermal
distribution. Here the convergence of the heat conductivity is related to decay rate of the 
autocorrelation function $c(\tau)$ as $\tau\rightarrow\infty$. The chain has normal conductivity
if the decay is fast enough for convergence of integral (\ref{f10}).

\begin{figure}[tb]
\includegraphics[angle=0, width=1\linewidth]{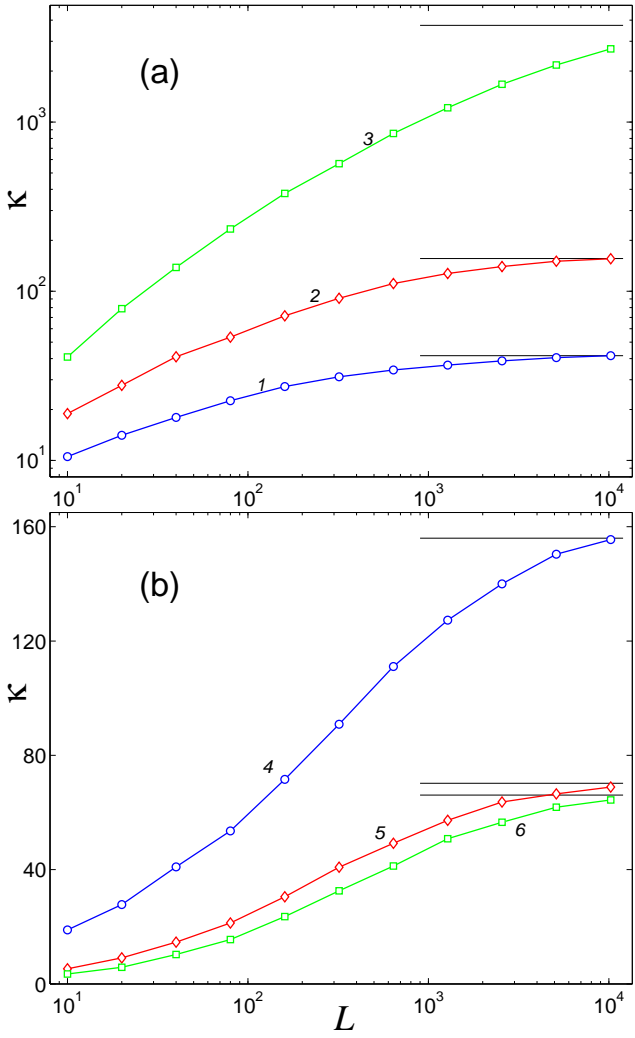}
\caption{(Color online)
Dependence of the heat conduction coefficient $\kappa$ on the length of chain $L$ for (a) temperatures $T=0.0001$,
0.001, 0.01 (curves 1, 2, 3) with packing density $p=1$, and (b)
for packing densities $p=1$, 10/11, 5/6 (curves 4, 5, 6) with temperature $T=0.001$.
Black straight lines are calculated using Green-Kubo formula (\ref{f9}).
}
\label{fig02}
\end{figure}

\begin{figure}[tb]
\includegraphics[angle=0, width=1\linewidth]{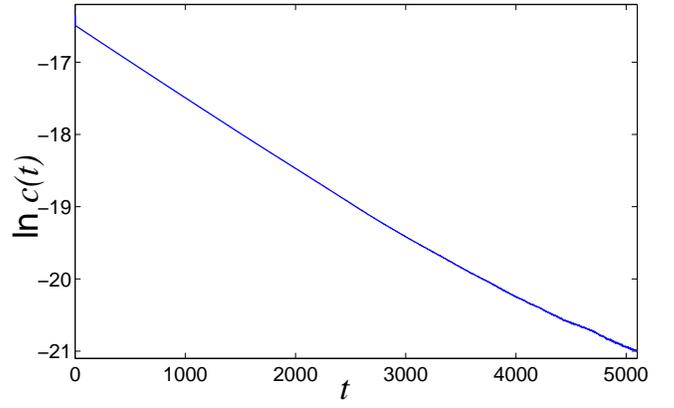}
\caption{
Exponential decay of the autocorellation function $c(\tau)$ for a
chain of elastic disks with  packing density  $p=10/11$ and temperature  $T=0.001$.
}
\label{fig03}
\end{figure}

The numerical simulation of the heat transport demonstrates convergent conduction in the chain
of elastic disks in all range of the temperatures and densities (see Fig. \ref{fig02}).
The convergence is also confirmed by the behavior of the autocorrelation function $c(\tau)$.
When $\tau\rightarrow\infty$ the function $c(\tau)$ decreases  exponentially,
i.e. behaves in leading order as  $\exp(-\lambda\tau)$,
$\lambda>0$ (see Fig. \ref{fig03}). Thus Green-Kubo formula  (\ref{f10}) implies finite conduction coefficient.
 Moreover, both methods (equilibrium and non-equilibrium modeling) yield
similar results for long chains (Fig. \ref{fig02}), which provides additional validation
of the simulation results.
%---------------------------- Fig. 2 ------------------------------------
%---------------------------- Fig. 2 ------------------------------------
%---------------------------- Fig. 3 ------------------------------------
%---------------------------- Fig. 3 ------------------------------------
%---------------------------- Fig. 4 ------------------------------------
\begin{figure}[tb]
\includegraphics[angle=0, width=1\linewidth]{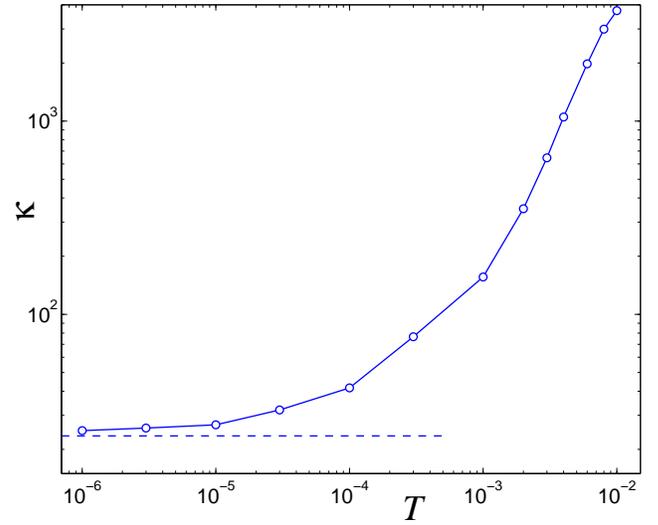}
\caption{
Dependence of the heat conduction coefficient $\kappa$ on the modeling temperature of the chain $T$
(the packing density is $p=1$). The dashed horizontal line represents the heat conductivity
at the limit $T\rightarrow 0$.
}
\label{fig04}
\end{figure}
%---------------------------- Fig. 4 ------------------------------------

Figure \ref{fig04} shows that heat conductivity coefficient increases monotonically as the
temperature increases. For $T\rightarrow 0$
the coefficient approaches a value  $\kappa_0>0$, which describes the heat conductivity of a chain
consisting of harmonic elastic disks. It is a kind of expected, since for small displacements the
interaction potential (\ref{f4}) can be replaced by harmonic potential of repulsion.

The heat conductivity of a chain with such semi-harmonic potential of interaction between
particles was investigated recently  \cite{GS14}. The finite conductivity of the chain is
obtained for all values of the packing density, (in particular for dense packing where $p=1$),
and the value of the heat conductivity coefficient is independent of temperature.
As $T\rightarrow\infty$ the heat conductivity of the chain sharply increases. The reason is that
for higher energies, the interaction between the disks is governed by the hard core of the disk
potential. It is well-known that a system of hard disks  is completely
integrable.
%---------------------------- Fig. 5 ------------------------------------
\begin{figure}[tbh]
\includegraphics[angle=0, width=1\linewidth]{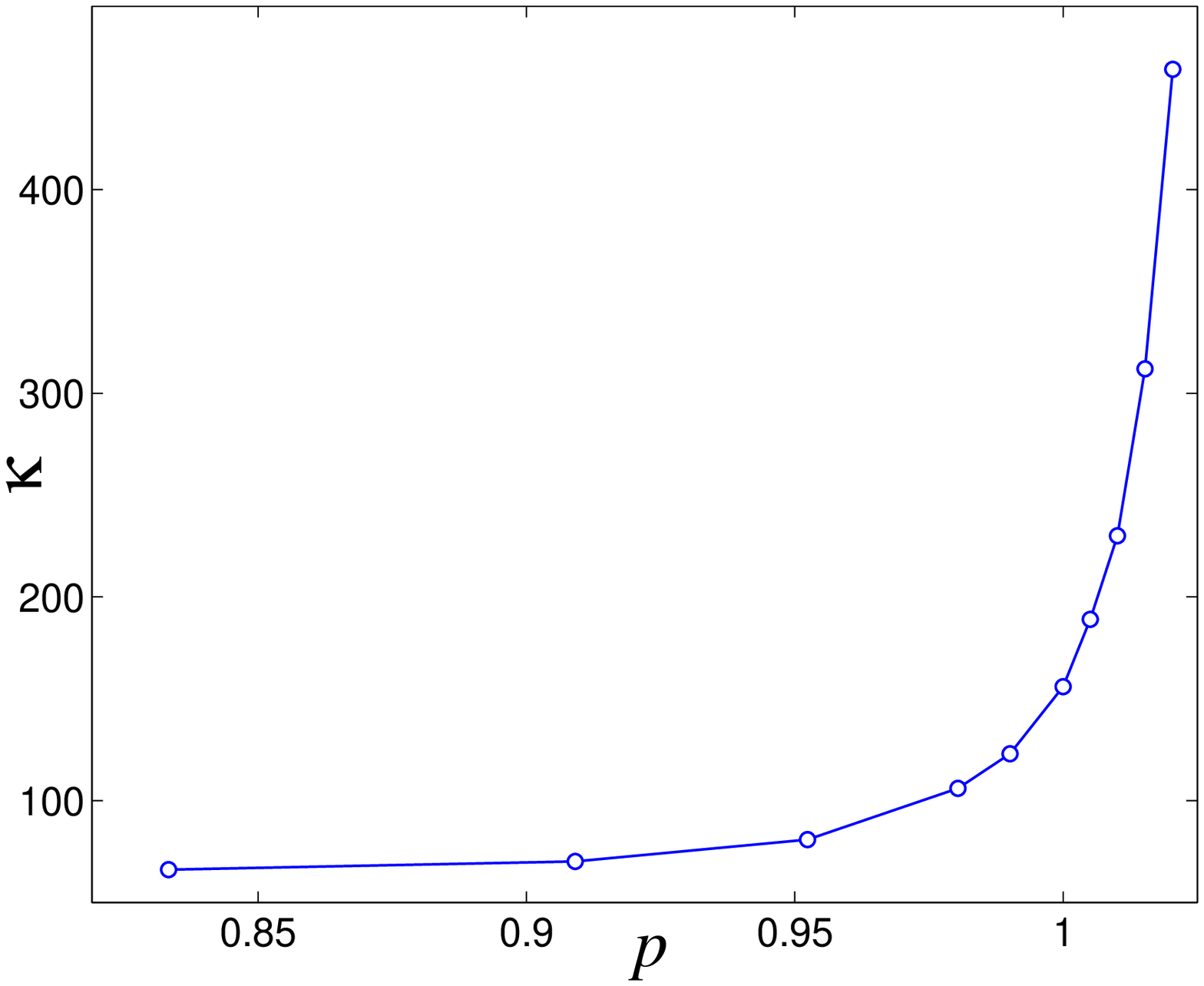}
\caption{
Dependence of the heat conduction coefficient $\kappa$ on the packing density of the chain, $p$
(the modeling temperature  is $T=0.001$).
}
\label{fig05}
\end{figure}
%---------------------------- Fig. 5 ------------------------------------

A limit case of elastic interaction can be obtained by increasing the density of the chain.
Maximal value of the chain density, $p_m=1/d$, is obtained when the hard cores of the
neighboring disks come into a contact. The dependence of the heat conduction coefficient on the
density of the chain is presented in Fig.~\ref{fig05}. The heat conductivity increases
monotonically with increase in density. For $p\rightarrow p_m$ the conductivity sharply increases, as one should expect.

\section{Heat conduction in quasi-1D chain}

Let us now consider a heat conduction problem of system of disks located in a long rectangular
channel: $0<x<L_x$, $0<y<L_y$, where $L_x$ and $L_y$ are the length and the width of the channel
respectively, and $(x,y)$ are coordinates of a center of a disk. Simulation of the system in the channel 
requires defining interactions of the disks with the channel walls.
We will assume that the walls are rigid, and the interaction is given by potential
\begin{equation}
U(x,y)=U_1(x)+U_2(y),
\label{f11}
\end{equation}
where
\begin{eqnarray}
U_i(u)&=&\frac18(1-d)^2\left(\frac{1/2-u}{u-d/2}\right)^2,\nonumber\\
&~& \mbox{for}~~d/2<u\le 1/2, \nonumber\\
U_i(u)&=& 0, ~~\mbox{for}~~ 1/2<u<L-1/2, \label{f12}\\
U_i(u)&=&\frac18(1-d)^2\left(\frac{L-1/2-u}{u-L+d/2}\right)^2,\nonumber\\
&~& \mbox{for}~~L-1/2\le u<L-d/2,\nonumber
\nonumber
\end{eqnarray}
for $i=1$ the potential defines the interaction with vertical walls: $u=x$, $L=L_x$,
for $i=2$ -- the interaction with horizontal walls: $u=y$, $L=L_y$.
As previously, we will use the smoothened form of this potential -- see appendix \ref{a2}.

If the width of the channel $L_y\le 2d$, then the neighbor disks cannot exchange their positions
along $x$ axis due to the hard cores. Consequently, in a narrow channel we obtain a
quasi-1D chain of disks.
If at the initial moment the system consists of $N$ disks with ascending
order of $x$-components of their centers [$0<x_1<...<x_{n-1}<x_n<...<x_N<L_x$ ($L_x>Nd$)],
then this order will always remain unchanged.

The dimensionless Hamiltonian of the chain is expressed as:
\begin{eqnarray}
{\cal H}&=&\sum_{n=1}^N\frac12({x'}_n^2+{y'}_n^2)
+\sum_{n=1}^{N-1}V(r_{n,n+1})+\sum_{n=1}^{N-2}V(r_{n,n+2}) \nonumber\\
&~& \sum_{n=1}^N U_2(y_n) +U_1(x_1)+U_1(x_N),\label{f13}
\end{eqnarray}
where the vector  ${\bf x}_n=(x_n,y_n)$ defines the coordinates of  $n$-th disk, and
$r_{i,j}=[(x_j-x_i)^2+(y_j-y_i)^2]^{1/2}$
is the distance between the centers of disks $i$ and $j$.

\begin{figure}[tb]
\includegraphics[angle=0, width=1\linewidth]{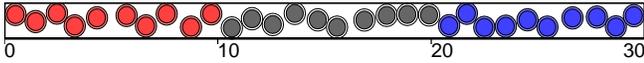}
\caption{(Color online)
Model of quasi-1D chain with length $L_x=30$, width $L_y=1.6$ and  with left end (red disks)
attached to $T=T_+$ thermostat and right end (blue disks) attached to $T=T_-$  thermostat
(density $p=1$).
}
\label{fig06}
\end{figure}

As previously we insert boundaries ($L_0=10$) into Langevin thermostats
with temperatures $T_\pm=(1\pm 0.05)T$. An example of  a quasi-1D channel is shown on Fig. \ref{fig06}.
%---------------------------- Fig. 6 ------------------------------------
%---------------------------- Fig. 6 ------------------------------------

Corresponding equations of motion are written as follows:
\begin{eqnarray}
{\bf x}''_n&=&-\partial {\cal H}/\partial {\bf x}_n -\gamma {\bf x}'_n+\Xi_n^+,~~\mbox{if}~~x_n< L_0,\nonumber\\
{\bf x}''_n&=&-\partial {\cal H}/\partial {\bf x}_n,~~\mbox{if}~~L_0\le x_n\le L_x-L_0,\label{f14}\\
{\bf x}''_n&=&-\partial {\cal H}/\partial {\bf x}_n -\gamma {\bf x}'_n+\Xi_n^-,~~\mbox{if}~~x_n> L_x-L_0,\nonumber
\end{eqnarray}
where $\Xi_n^\pm=(\xi_{n,1},\xi_{n,2})$ is a Gaussian white noise which models the interaction
with the thermostat, and is normalized by the conditions $\langle\xi_{n,i}^\pm(\tau)\rangle=0$,
$\langle\xi_{n,i}^+(\tau_1)\xi^-_{k,j}(\tau_2)\rangle=0$,
$\langle\xi_{n,i}^\pm(\tau_1)\xi_{k,j}^\pm(\tau_2)\rangle=2\gamma T_\pm\delta_{nk}\delta_{ij}\delta(\tau_2-\tau_1)$.

Verlet Velocity method was used in order to obtain the numerical solution of (\ref{f14}).
The following initial configuration of the chain was considered:
\begin{eqnarray}
x_n(0)=(n-1)a,~x'_n(0)=0,\nonumber\\
y_n(0)=[1+(L_y-1)(1+(-1)^n)]/2,~y'_n(0)=0,\nonumber
\end{eqnarray}

The thermal equilibrium between the chain and the thermostats has been reached and is manifested
by a stationary heat flux, $J$ and the local temperature distribution $T(x)$.

We will, as previously, calculate the local temperature distribution of the chain in terms of
distribution of the kinetic energy of the disks. The rectangular channel of length $L_x$ which
consists of  $N$ disks, is divided into unit-length cells $i-1\le x<i$, $i=1,...,L_x$.
Then the average number of disks in $i$-th cell is $\bar{n}_i$,  $\bar{E}_i$ is the average
kinetic energy in the cell. The temperature of the cell is given by
$T(i)=\bar{E}_i/\bar{n}_i$.

The energy transfer from disk $n$ to the neighbor disk
$n+1$ is given by $J_n=\langle j_n\rangle_\tau$, where
$$
j_n=x_{n,1}'h_n-\sum_{k=1}^2 (x_{n+k,1}-x_{n,1})({\bf x}'_n,{\bf F}({\bf x}_{n},{\bf x}_{n+k})),
$$
and the vector
$$
{\bf F}({\bf x}_1,{\bf x}_2)=\left(\frac{\partial V(r_{1,2})}{\partial x_{2,1}},
\frac{\partial V(r_{1,2})}{\partial x_{2,1}}\right),~~r_{1,2}=|{\bf x}_2-{\bf x}_1|,
$$
the energy density distribution along the system:
$$
h_n=\frac12\{{x'_{n,1}}^2+{x'_{n,2}}^2+\sum_{k=1}^2[V(r_{n-k,n})+V(r_{n,n+k})]\}.
$$
(see \cite{LLP03}).

The total heat flux $J$
was calculated as the mean value of the work done by
the end thermostats -- see Eq. (\ref{f7}).
%---------------------------- Fig. 7 ------------------------------------
\begin{figure}[tb]
\includegraphics[angle=0, width=1\linewidth]{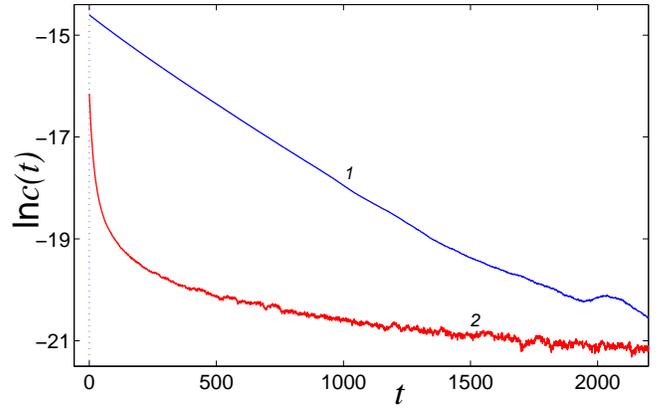}
\caption{
Exponential decay of the autocorellation function  $c(\tau)$ for a
system of elastic disks. The packing density is $p=1$ and the modeling
temperature $T=0.001$, width of the channel $L_y$=1 and 1.5 (curves 1 and 2).
}
\label{fig07}
\end{figure}
%---------------------------- Fig. 7 ------------------------------------
%---------------------------- Fig. 8 ------------------------------------
\begin{figure}[tb]
\includegraphics[angle=0, width=1\linewidth]{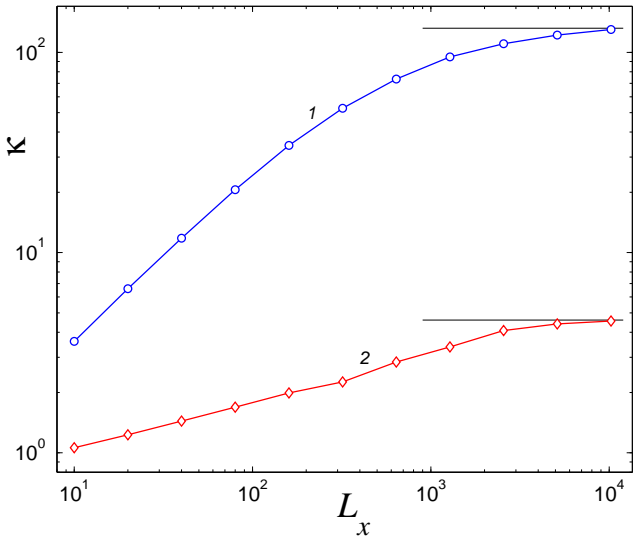}
\caption{(Color online)
Dependence of the heat conduction coefficient $\kappa$ on the length of the quasi-1D channel
$L_x$ for temperature $T=0.001$,
packing density $p=1$, and width of the channel $L_y=1$ and 1.5  (curves 1 and 2).
Black straight lines are calculated using Green-Kubo formula (\ref{f10}).
}
\label{fig08}
\end{figure}
%---------------------------- Fig. 8 ------------------------------------
%---------------------------- Fig. 9 ------------------------------------
\begin{figure}[tb]
\includegraphics[angle=0, width=1\linewidth]{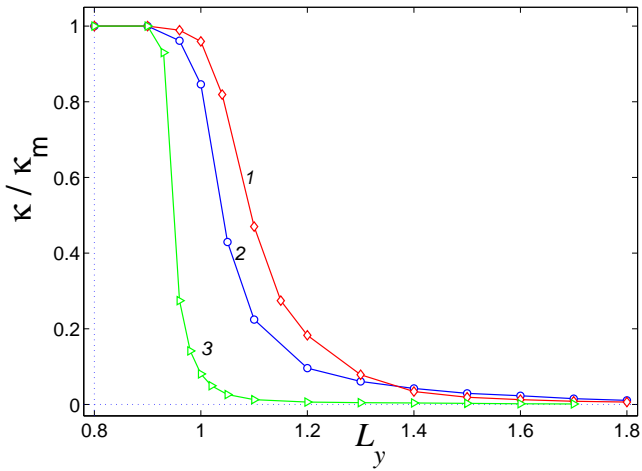}
\caption{(Color online)
Dependence of the normalized heat conduction coefficient $\kappa/\kappa_{m}$
on the width of the quasi-1D channel $L_y$
for temperature $T=0.0001$ ($\kappa_{m}=47$), $T=0.001$ ($\kappa_{m}=156$) and
$T=0.01$ ($\kappa_{m}=3728$) -- curves 1, 2 and 3.
}
\label{fig09}
\end{figure}
%---------------------------- Fig. 9 ------------------------------------

The heat conductivity was calculated from direct modeling
of heat transfer using Eq. (\ref{f8}), and also from  Green-Kubo formula (\ref{f10}).
The exponential decay of the autocorrelation function $c(\tau)$ (see Fig.~\ref{fig07}) provides
the convergence of the integral in Green-Kubo formula (\ref{f10}). In Fig.~\ref{fig08}
we depict the heat conductivity versus length of the channel $L_x$ for quasi-1D chain
($L_y=1$ and $L_y=1.5$). The heat conductivity of the chain saturates in the thermodynamic limit
and the results are validated by Green-Kubo formula (straight black lines).

In the considered 1D chain the scattering occurs only as a result of multiple collisions
\cite{GS14}. Narrow two-dimensional channel supplies an additional
scattering mechanism -- the energy partially transfers from longitudinal  to transversal
components of motion. This effect becomes more significant if the width of the channel
increases. Therefore, we may expect the thermal conductivity of disks chain
to decrease as the channel becomes wider. The numerical calculations validate this expectation,
the thermal conductivity coefficient monotonically decreases with increased width of the channel
-- see Fig.~\ref{fig09}. The decrease is very significant in the width diapason $1< L_y<1.2$.
 As we see from Fig. \ref{fig09}, the decrease in the conductivity
is more sharp when the temperature increases. The presence of the
hard cores leads to an increase in stiffness of the collisions as
the temperature grows. As a result, the time span of  the individual collision and the probability of triple collisions
decrease, and the scattering occurs primarily due to two-dimensionality.

According the results of the numerical modeling we may conclude, that
the transfer from one-dimensional dynamics to two-dimensional occurs
in the width $1< L_y<1.2$. For width $L_y>1.2$ the 
decrease in the heat conductivity primarily stems from the 2D effects.

\section{Quasi-1D Billiard}

In order to understand the effect of the "additional dimensions"\  on thermal conductivity of the
quasi-1D chains, it is instructive to consider the case of hard disks,
which corresponds to $d=1$ (the diameter of the hard core equals to the diameter of the disk).
In this case all collisions occur instantly and are strictly pairwise. The only scattering
mechanism is the exchange of energy between vertical and horizontal components of momentum
of colliding disks. We should notice that a 1D chain of hard disks is completely
integrable system. In 2D system the picture changes -- in this situation collisions lead
to an appearance of chaotic dynamics.

We consider a chain of disks in a narrow channel with width $L<2d$. Such a width prevents the
possibility of disks with diameter $d$ to exchange positions, and we deal with the quasi-1D chain.
However, vertical displacements of disks in this model will enable the scattering.

Firstly we examine the direct modeling of heat transfer. The 2D rectangular channel has the width
$L_y$ and the length $L_x$: $0<x<L_x$, $0<y<L_y$,
$L_y/L_x\ll1$. The collisions of the disks with the walls
of the channel are elastic.

The ends of the channel are attached to thermostats with temperature $T_+$  at the left end
and $T_-$  at the right end of the channel. In order to thermalize the chain, we use hot boundary
ends thermostat. At the moment of the collision of a disk with the left end wall its
horizontal coordinate is $x=0.5$. Right after the collision the horizontal component of disk's
velocity is $v_x>0$, which value is defined according to Maxwell distribution
$P(v)=(|v|/T)\exp(-v^2/2T)$  with $T=T_+$. The vertical component of disk's velocity remains
unchanged  after the collision. At the moment of the collision of a disk with the right end
wall its horizontal coordinate is $x=L_x-0.5$. Right after the collision the sign of the
horizontal component of disk's velocity is negative, $v_x<0$, and the value is defined according
to Maxwell distribution $P(v)=(|v|/T)\exp(-v^2/2T)$  with $T=T_-$. If a disk is located at
the left end of the channel, i.e $0.5<x<L_0$, and it collides with a top(bottom) wall,
at the moment of such a collision its horizontal coordinate is $y=0.5$ or $y=L_y-0.5$,
the disk will change the sign of its vertical component of the velocity, $v_y$,
and the value of this component is calculated according to Maxwell
distribution for $T=T_+$ ($v_x$ remains unchanged). The same approach is used in order
to account the collisions with top and the bottom
at the right and of the channel ($L_x-L_0<x<L_x-0.5$), where $T=T_-$.

In such a model, the interaction of the disks with thermostats occurs only through the collisions with boundary
walls. We then compute the work done by the thermostats. If before the collision with a hot
wall the velocity of the disk was $(v_x(t_i-0),v_y(t_i-0))$, and after the collision
--  $(v_x(t_i+0),v_y(t_i+0))$, then at the moment of the collision,  $t=t_i$ the work done
by the thermostat is  $\Delta E(t_i)=E(t_i+0)-E(t_i-0)$, where $E(t)=[v_x^2(t)+v_y^2(t)]/2$ --
the kinetic energy of the disk. If in the time interval  $[0,t]$ there occur $N_t$ collisions with the
thermostat walls (the sequence of collision times $\{ t_i\}_{i=1}^{N_t}\in [0,t]$),
then the average work of the thermostat is
$$
j_\pm(t)=\frac1t \sum_{i=1}^{N_t}\Delta E_\pm(t_i)
$$
(sign plus for left thermostat and minus for right thermostat).
The average intensity of the work is $J_\pm=\lim_{t\rightarrow\infty}j_\pm(t)$.

Let $p$ be the linear density of disks in the channel (number of disks $N=(L_x-1)p$).
We consider a system of disks with linear density $p=1$, then $L_x=N+1$.
The initial configuration of the system is determined as follows:
\begin{eqnarray}
&&x_i(0)=1+(i-1)a_x,~~y_i(0)=0.5+(L_y-1)\zeta_i,\nonumber \\
&&x_i'(0)=v_{i,1},~~y_i'(0)=v_{i,2},~~i=1,2,...,N \nonumber
\end{eqnarray}
where $a_x=(L_x-1)/N$ is  period of the chain (for density $p=1$  the period $a_x=1$),
$\zeta_i$  are random numbers which are distributed uniformly in the segment  [0,1],
$v_{i,1}$ and $v_{i,2}$ are random velocities with distribution $P(v)=\exp[-v^2/2T]/\sqrt{2\pi T}$,
temperature  $T=(T_++T_-)/2$, $T_+$ and $T_-$ are temperatures of the left and the right
thermostats respectively.
%---------------------------- Fig. 10 ------------------------------------
\begin{figure}[tbh]
\includegraphics[angle=0, width=1\linewidth]{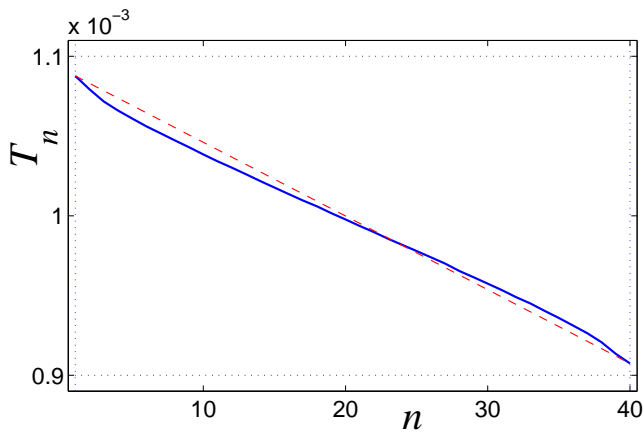}
\caption{(Color online)
Local distribution of temperature $T_n$ in the chain of hard disks in a 2D channel with length
$L_x=41$ and width $L_y=1.5$ (packing density $p=1$, number of disks $N=(L_x-1)p=40$).
}
\label{fig10}
\end{figure}
%---------------------------- Fig. 10 ------------------------------------

We will use the following numerical values in order to simulate the dynamics of the system
$L_x=N+1$, $L_y=1.5$, 1.8, $L_0=1.5$, $T_+=0.0011$, $T_-=0.0009$. Average values of heat
fluxes $J_+$, $J_-$ are calculated after formation of the steady heat flux along the chain
(in the system with steady heat flux  $J=J_+=-J_-$).
The temperature distribution in the chain is defined as $T_n=\langle x_n'(t)^2+y_n'(t)^2\rangle_t/2$.

Numerical modeling of the dynamics of the system reveals the linear temperature gradient in
2D channel, see Fig.~\ref{fig10}. It apparently appears due to violation of integrability by 
to the non-central collisions. At both ends of the chain, where the interaction with the thermostats
takes place, we obtain heat resistance, due to which the temperature of the left end is
always lower than the temperature of the left wall $T_1<T_+$, and the temperature of the right end
is always higher than the temperature of the right wall $T_N>T_-$  (this end effect disappears
with increase of the length of the chain). In order to account for this effect, we calculate the
thermal conductivity coefficient in terms of temperature difference:
\begin{equation}
\kappa(L_x)=J(L_x-1)/(T_1-T_N).
\label{f15}
\end{equation}
%---------------------------- Fig. 11 ------------------------------------
\begin{figure}[tb]
\includegraphics[angle=0, width=1\linewidth]{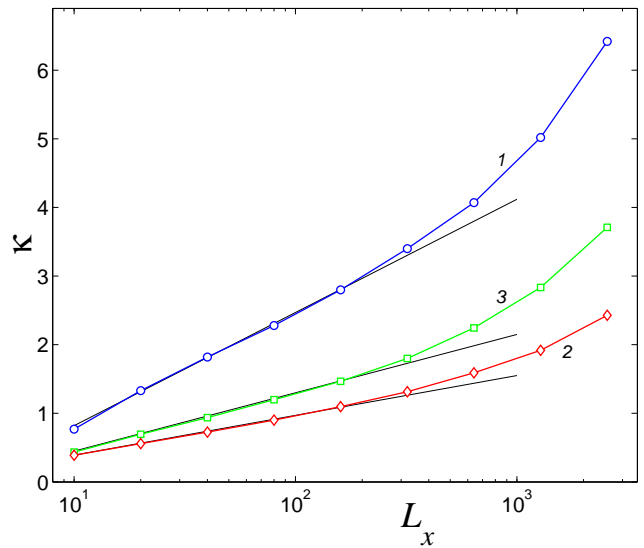}
\caption{(Color online)
Dependence of the heat conduction coefficient $\kappa$ on length of the channel $L_x$
for chain of hard disks (curve 1 and 2) and spheres (curve 3).
The width of the channel $L_y=1.5$ (curves 1 and 3) and 1.8 (curves 2).
Black straight lines correspond to the
logarithmic relations $\kappa=\alpha\ln L_x$ for $\alpha=0.25$, 0.37 and 0.72.
}
\label{fig11}
\end{figure}
%---------------------------- Fig. 11 ------------------------------------

The dependence of the thermal conductivity coefficient $\kappa$  on the length of the channel
$L_x$ is presented in Fig.~\ref{fig11}. According to the figure, the conductivity of the chain
grows monotonically with the length of the chain. For lengths  $L_x<300$  the increase in
conductivity is logarithmic $\kappa\sim \log(L_x)$, however, for large values of $L_x$ we observe
an increase in the growth rate of conductivity. In other terms, one observes the effective crossover from "genuine"\
2D behavior for relatively small $L_x$ , characterized by logarithmic divergence of the heat
conduction coefficient, to more fast "quasi-1D"\  divergence for longer lattices.
As one could expect from previous sections, the rate of the growth decreases with an
increase of the width of the channel, but qualitatively the behavior remains the same.

Numerical simulation of the heat conduction shows that the chain of  hard disks in the
narrow two-dimensional channel has divergent heat conductivity, as expected \cite{DN03,LL07,MT13}.
In order to further validate this result we examine the behavior of the autocorellation function.
In the case of colliding billiard particles, computation of this function requires certain
modification as compared to more common cases. We consider 2D channel with length $L_x$
under periodic boundary conditions in the horizontal direction. The number of disks in this
channel is $N=L_xp$ (the linear density is defined as unity, $p=1$, so the number of the disks
equals the length of the chain). The disks are initiated with normally distributed random
velocities, so that the complete chain is thermalized with the temperature $T=0.001$.

In a narrow channel with $L_y<1+\sqrt{3/4}$ the collisions can occur only between the neighbor disks.
Let us consider that at the moment $t=t_i$ we indicate the collision between the disks $n_i$ and $n_i+1$.
If before the collision the velocity of the disk $n_i$ was $(v_{n_i,x}(t_i-0),v_{n_i,y}(t_i-0))$,
and after the collision  $(v_{n_i,x}(t_i+0),v_{n_i,y}(t_i+0))$, then the collision leads to the
change in kinetic energy
$
\Delta E_{n_i}=[v_{n_i,x}^2(t_i+0)+v_{n_i,y}^2(t_i+0)-v_{n_i,x}^2(t_i-0)-v_{n_i,y}^2(t_i-0)]/2.
$
Thus, the collision leads to transfer of energy  $-\Delta E_{n_i}$ from particle $n$ to particle $n+1$.
If we take time increment $\Delta t$, which is much larger than an average time between
collisions, we can determine the value of heat flux from node $n$ to node $n+1$ in time interval
$[t,t+\Delta t]$:
$$
j_n(t)=-\frac{1}{\Delta t}\sum_{i:t_i\in [t,t+\Delta t]} \Delta E_{n_i}.
$$

After calculating the time-dependent total heat flux  $J_s(t)=\sum_{n=1}^Nj_n(t)$
we can find the autocorrelation function
$c(\tau)=\langle J_s(t)J_s(t-\tau)\rangle_t$.
%---------------------------- Fig. 12 ------------------------------------
\begin{figure}[tbh]
\includegraphics[angle=0, width=1\linewidth]{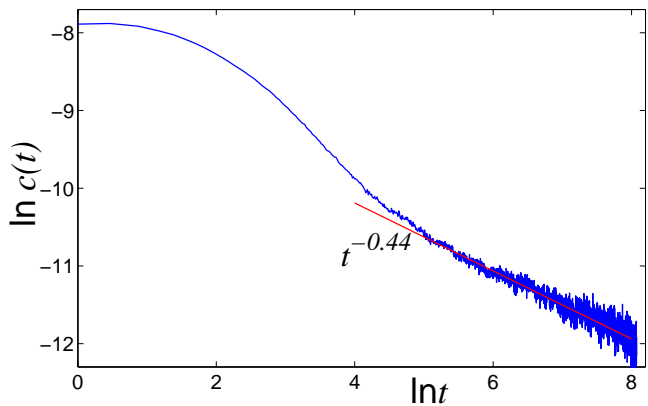}
\caption{(Color online)
Power law decay of the autocorrelation function $c(t)$ for chain of  hard disks
in 2D channel with width $L_y=1.8$ and temperature $T=0.001$. The straight line corresponds
to the power function  $t^{-0.44}$.
}
\label{fig12}
\end{figure}
%---------------------------- Fig. 12 ------------------------------------

We consider a system of $N=10^4$ disks in order to compute the autocorrelation function $c(\tau)$.
The temperature of the system is $T=0.001$ and the time-increment is $\Delta t=0.8$.
The dynamics of the thermalized system was observed in time interval $0\le t\le 20000\Delta t$.

The behavior of the autocorrelation function $c(t)$ for the chain of disks in a channel with width
$L_y=1.8$ is presented in Fig.~\ref{fig12}. The Figure shows that for times $t>150$ the
autocorrelation function decays as according to power law:
$c(t)\sim t^{-\alpha},~~t\rightarrow\infty$, where $\alpha=0.44<1$.
Such a behavior leads to divergence of heat conduction, which validates the results revealed
in direct modeling of heat transfer.

\section{Quasi-1D Billiard in three dimensions.}

Now we consider a chain of hard three-dimensional spheres located in a rectangular
channel $0<x<L_x$, $0<y<L_y$, $0<z<L_y$, where $L_x$ is the length and $L_y$ is the width.
The analysis methods for heat conduction in the system of two-dimensional disks are generalized
in the case of chain of 3D spheres.

Figure \ref{fig11} shows that the addition in dimensionality leads to an increase in the scattering
of kinetic energy, and, as a result, the thermal conductivity coefficient $\kappa(L_x)$
decreases. However, the conductivity continues to grow monotonically with an increase in the
length of the system. At low values of $L_x$  the growth is logarithmic  $\kappa\sim \log(L_x)$.
The rate of the growth increases as the length $L_x$ is increased. This implies that the
conductivity of the chain of 3D spheres in a narrow channel diverges. The divergence is also
validated by the behavior of the autocorrelation function.

\section{Control of the heat flux in  narrow channels}

If the width of a narrow rectangular channel is increased, the "two-dimensionality"\
is more pronounced. It follows, that an increase in the width of the
channel should lead  to the decrease of the heat flux along the channel. Due
to this fact we may significantly change the value of the heat flux
 by changing the width of the channel. We will demonstrate this on
 a system of hard disks confined in a two-dimensional
 rectangular channel.
%---------------------------- Fig. 13 ------------------------------------
\begin{figure}[tb]
\includegraphics[angle=0, width=1\linewidth]{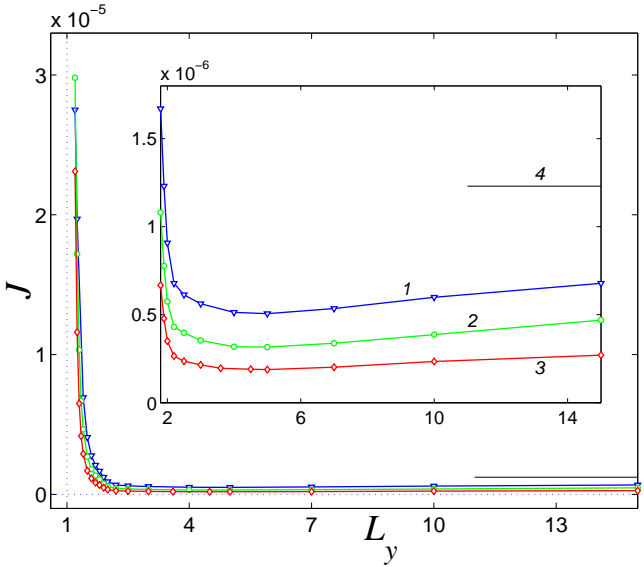}
\caption{(Color online)
Dependence of the heat flux $J$ on the width of channel $L_y$
in a rectangular channel of size $L_x\times L_y$,
filled with hard disks (number of disks corresponds to the length of the channel)
for lengths $L_x=43$, 83, 163 (curves 1, 2, 3). Temperature of boundary walls $T_+=0.00105$
and $T_-=0.00095$. The straight line 4 corresponds to the limit value of heat flux when
$L_y\rightarrow\infty$.
}
\label{fig13}
\end{figure}
%---------------------------- Fig. 13 ------------------------------------

We consider a channel with length $L_x$ and width $L_y$.
We locate a chain of hard disks of diameter $d=1$ within the channel, so that
the linear density of the chain is unit. For this sake it is enough to define the
number of disks $N$ to be equal to the length of the channel $L_x$. The heat transfer is modeled
using hot boundary ends thermostat with the left wall temperature $T_+=0.00105$
and right wall  temperature $T_-=0.00095$. The length of the channel remains fixed and
we examine the heat flux along the channel $J$ as a function of width of the channel $L_y$.

For the width of the channel $L_y=d=1$ we obtain a one-dimensional chain of densely packed
hard disks. Here the momentum instantly passes from one wall to the opposite, so the heat flux is infinite.
If we increase the width of the channel, the disks are
able to displace and to move in both horizontal and vertical directions. Due to the displacements
and collisions of the disks we obtain a finite stationary heat flux along the channel.
The dependence of the heat flux $J$ on the width of the channel $L_y$ is shown on Fig.~\ref{fig13}
for the values $L_x=43$, 83, 163. It can be inferred from the figure that the increase in the width
of the narrow channel leads to a sharp decrease of the heat flux. The wider the channel,
the stronger effect of "two-dimensionality"\  on the dynamics is observed.
The minimal value of the heat flux is obtained for $L_y=4\div 5$. Further increase of the width
leads to slow growth of heat flux. Heat flux monotonically approaches a limit value
$J\nearrow 1.23\cdot 10^{-6}$ for $L_y\nearrow\infty$. At this limit the heat flux remains almost unaffected by the collisions due to relatively small particle density. 
For fixed horizontal length of the channel the value
of the heat flux depends only on the temperature difference of the boundary walls.

\section{Discussion and concluding remarks}
Significance of low-dimensional models for physical applications is often questioned,
especially as they demonstrate a behavior different from their three-dimensional counterparts.
Indeed, every real system is three-dimensional. In the same time, the results presented above
indicate that in conditions of confinement and large enough aspect ratio two- and
even three-dimensional systems demonstrate clear features of quasi-one-dimensional behavior.

Crossover to this quasi-one-dimensional behavior requires further exploration.
In this paper, we observe two different scenarios. The transition can be rather sharp
or smooth crossover from logarithmic to power-like divergence in the case of "billiard"\ model.
One can conjecture that this lack of universality is related to convergence or divergence
of the heat conduction coefficient in the thermodynamic limit; this issue might be a subject
of further investigation.

\section{Acknowledgments}
The authors are very grateful to Israel Science Foundation (grant 838/13) and to Lady Davis
Fellowship Trust for financial support of their work. A.V.S. is grateful to the Joint
Supercomputer Center of the Russian Academy of Sciences for the use of computer facilities.

\appendix
\section{Smoothening of  interaction potential of disks.}\label{a1}
Using of piecewise analytic potentials in numerical simulations may
lead to fast accumulation of errors in numerical integration. In order
to prevent such errors we implement local "smoothening" of potentials
in neighbourhood of points where derivatives are not continuous.

Thus we approximate potential $V(r)$ by smoothened potential in the form
\begin{equation}
V_h(r)=\frac14\left[\sqrt{V(r)+hf(r)}+\mbox{sgn}(1-r)\sqrt{V(r)}\right]^2, \label{fa1}
\end{equation}
where the value parameter $h>0$ determines the accuracy of the smoothening, $f(r)$ is
a positive function localized in the neighborhood of $r=1$.
At the limit  $h\rightarrow 0$ the smoothened potential
$V_h(r)\rightarrow V(r)$ for $r<1$ and $V_h(r)\rightarrow 0$ for $r\ge 1$.
%---------------------------- Fig. a1 ------------------------------------
\begin{figure}[tbh]
\includegraphics[angle=0, width=1\linewidth]{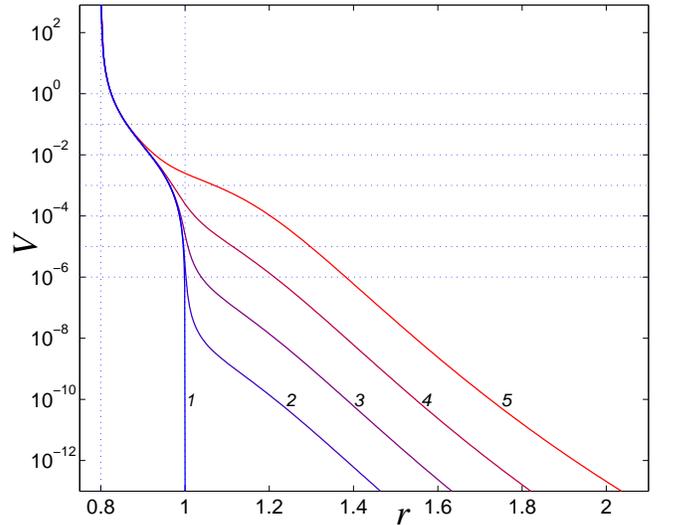}
\caption{(Color online)
Repulsive interaction potential  $V(r)$ (\ref{f4}) (curve 1, parameter $d=0.8$)
and smoothened potential  $V_h(r)$ (\ref{fa2}) for $h=0.00001$, 0.0001, 0.001, 0.01
(curves 2, 3, 4 ,5).
}
\label{figa1}
\end{figure}
%---------------------------- Fig. a1 ------------------------------------
%---------------------------- Fig. a2 ------------------------------------
\begin{figure}[tbh]
\includegraphics[angle=0, width=1\linewidth]{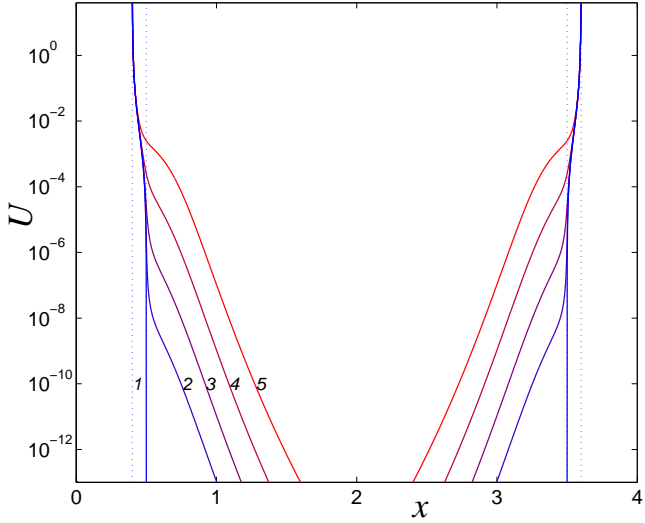}
\caption{(Color online)
Interaction potential with the walls $U(u)$ (\ref{f12}) (curve 1)
and smoothened potential $U_h(u)$ for the length between walls  $L=4$
and smoothening parameter $h=0.00001$, 0.0001, 0.001, 0.01
(curves 2, 3, 4 ,5).
}
\label{figa2}
\end{figure}
%---------------------------- Fig. a2 ------------------------------------

For numerical considerations we adopt a simple form of the smoothening function
$f(r)=[1+5(r-1)^2]^{-6}$, which allows to use a single expression for modeling the
repulsive interaction between disks:
\begin{equation}
V_h(r)=c_1\left[\sqrt{\left(\frac{1-r}{r-d}\right)^2+hc_2f(r)}+\frac{1-r}{r-d}\right]^2,
\label{fa2}
\end{equation}
with the coefficients  $c_1=(1-d)^2/8$, $c_2=2/(1-d)^2$.
The smoothened potential (\ref{fa2}) is presented in Fig.~\ref{figa1}.

In order to preclude the smoothening artifacts, the accuracy $h$ should correlate with
the temperature $T$ of the chain. As we see from Fig.~\ref{figa1} for $T\ge 0.5$  it is
enough to adopt $h=0.01$, for $0.05\le T<0.5$ -- $h=0.001$, for $0.005\le T<0.05$ -- $h=0.0001$
and for $T<0.005$ -- $h=0.00001$.

\section{Smoothening of  interaction potential of disk with walls.}\label{a2}
In order to avoid numeric complications we approximate potential (\ref{f12}) $U_i(r)$ $(i=1,2)$
by smoothened potential in the form
\begin{eqnarray}
U_h(u)=c_1\left\{\left[\frac{(c_3-u)^2}{(u-c_4)^2}+hc_2f(u-c_3)\right]^{1/2}+\frac{c_3-u}{u-c_4}\right\}^2\nonumber\\
+c_1\left\{\left[\frac{(c_5-u)^2}{(u-c_6)^2}+hc_2f(u-c_5)\right]^{1/2}+\frac{c_5-u}{u-c_6}\right\}^2,~~~~~
\label{fa3}\\
 c_1=(1-d)^2/32,~~c_2=8/(1-d)^2,~~c_3=1/2,~~c_4=d/2,\nonumber\\
 c_5=L-1/2,~~ c_6=L-d/2, \nonumber
\end{eqnarray}
where localized positive function $f(u)=[1+5u^2]^{-6}$.
Value parameter $h>0$ determines the accuracy of the smoothening.

The smoothened potential (\ref{fa3}) is presented in Fig.~\ref{figa2}.
The accuracy $h$ should correlate with
the temperature $T$ of the chain. As we see from Fig.~\ref{figa2} for $T\ge 0.5$  it is
enough to adopt $h=0.01$, for $0.05\le T<0.5$ -- $h=0.001$, for $0.005\le T<0.05$ -- $h=0.0001$
and for $T<0.005$ -- $h=0.00001$.

\end{document}